\documentclass{emulateapj}
\usepackage{apjfonts}
\usepackage{subfigure}
\usepackage{graphicx}

\newcommand{\Msun}      {\mbox{$\rm\,M_{\mathord\odot}$}}

\begin{document}

\lefthead{Identifications of Five {\em INTEGRAL} Sources via Optical Spectroscopy}
\righthead{Butler et al.}

\submitted{Submitted 2008 November 8; accepted 2009 March 3}
\title{Identifications of Five {\em INTEGRAL} Sources via Optical Spectroscopy}

\author{Suzanne C. Butler\altaffilmark{1,2},
John A. Tomsick\altaffilmark{2},
Sylvain Chaty\altaffilmark{3},
Juan A. Zurita Heras\altaffilmark{3},
Jerome Rodriguez\altaffilmark{3},
Roland Walter\altaffilmark{4},
Philip Kaaret\altaffilmark{5},
Emrah Kalemci\altaffilmark{6},
Mehtap \"Ozbey\altaffilmark{7}}

\altaffiltext{1}{Department of Astronomy, University of 
California, Berkeley, CA 94720-3411, USA
(email: sbutler@astro.berkeley.edu)}

\altaffiltext{2}{Space Sciences Laboratory, 7 Gauss Way, 
University of California, Berkeley, CA 94720-7450, USA
(e-mail: jtomsick@ssl.berkeley.edu)}

\altaffiltext{3}{AIM - Astrophysique Interactions Multi-\'echelles
(UMR 7158 CEA/CNRS/Universit\'e Paris 7 Denis Diderot),
CEA Saclay, DSM/IRFU/Service d'Astrophysique, B\^at. 709,
L'Orme des Merisiers, FR-91 191 Gif-sur-Yvette Cedex, France}

\altaffiltext{4}{INTEGRAL Science Data Centre, Observatoire
de Gen\`eve, Universit\'e de Gen\`eve, Chemin d'Ecogia, 16, 
1290 Versoix, Switzerland}

\altaffiltext{5}{Department of Physics and Astronomy, University of
Iowa, Iowa City, IA 52242, USA}

\altaffiltext{6}{Sabanc\i\ University, Orhanl\i -Tuzla, \.Istanbul,
34956, Turkey}

\altaffiltext{7}{Physics Department, Middle East Technical University, 06531, Ankara, Turkey}

\begin{abstract}
The International Gamma-Ray Astrophysics Laboratory ({\em INTEGRAL}) is discovering hundreds of new hard X-ray sources, many of which remain unidentified.  We report on optical spectroscopy of five such sources for which X-ray observations at lower energies ($\sim$0.5--10 keV) and higher angular resolutions than {\em INTEGRAL} have allowed for unique optical counterparts to be located.  We find that IGR J16426+6536 and IGR J22292+6647 are Type 1 Seyfert active galactic nuclei (with IGR J16426+6536 further classified as a Seyfert 1.5) which have redshifts of $z=0.323$ and $z=0.113$, respectively.   IGR J18308--1232 is identified as a cataclysmic variable (CV), and we confirm a previous identification of IGR J19267+1325 as a magnetic CV.  IGR J18214--1318 is identified as an obscured high mass X-ray binary (HMXB), which are systems thought to have a compact object embedded in the stellar wind of a massive star.  We combine {\em Chandra} fluxes with distances based on the optical observations to calculate X-ray luminosities of the HMXB and CVs, finding $L_{\rm 0.3-10~keV}=5 \times 10^{36} {\rm ~erg~s^{-1}}$ for IGR J18214--1318, $L_{\rm 0.3-10~keV}=1.3 \times 10^{32}{\rm ~erg~s^{-1}}$ for IGR J18308--1232, and $L_{\rm 0.3-10~keV}=6.7 \times 10^{32}{\rm ~erg~s^{-1}}$ for IGR J19267+1325.
\end{abstract}


\section{Introduction}
Since its launch on October 17, 2002, the International Gamma-Ray Astrophysics Laboratory ({\em INTEGRAL}) has discovered hundreds of new hard X-ray sources.  According to the most recent census, of the $\sim$500 sources detected in hard X-rays at energies greater than 20 keV, 214 had not been well-studied (or even detected in most cases) before \citep{bodaghee07}.  These new sources are called ``IGR'' sources, for {\em INTEGRAL} Gamma-Ray sources.  Of these IGR sources, 50 had been identified as active galactic nuclei (AGN), 32 as high-mass X-ray binaries (HMXBs), 6 as low-mass X-ray binaries (LMXBs), and 15 as sources such as cataclysmic variables (CVs), supernova remnants, and anomalous X-ray pulsars, leaving 111 unclassified.  Since this last census, much work has been done to identify the $\sim$50\% of unclassified IGR sources (corresponding to $\sim$25\% of all sources detected by {\em INTEGRAL}).  Many have been identified as AGN, LMXBs and HMXBs, as well as relatively rare systems such as heavily absorbed (i.e., ``obscured") supergiant HMXBs, supergiant fast X-ray transients, and Intermediate Polar (IP) CVs \citep[e.g.,][]{masetti09,chaty08}.  

{\em INTEGRAL} is particularly well suited for finding such systems.  Part of {\em INTEGRAL}'s Core Program involved scans of the Galactic plane \citep{winkler03}.  One expects to find HMXBs here, as they consist of a neutron star or black hole accreting from a short-lived massive star, which would not be expected to move far from its birthplace in star forming regions of the Galactic plane.  Other instruments observing at lower energies could not easily find these because of the obscuring dust and gas in the plane.  Furthermore, objects such as obscured HMXBs, which consist of a compact object embedded in the stellar wind of a supergiant star, suffer obscuration not only from the intervening medium, but also from local absorption \citep[e.g.,][]{rodriguez03,walter06}.  In addition, {\em INTEGRAL} has found disproportionately many IP CVs according to the previously known population statistics, since they emit at these high energies more than non-magnetic CVs or polar CVs \citep{barlow06}.  Currently, less than 10\% of known CVs are magnetic.  The majority of these are polars, so called since they show polarization of optical flux.  These are systems in which the white dwarf has a magnetic field strong enough to synchronize the orbital period of the binary with the spin of the white dwarf.  Intermediate polars have a weaker magnetic field that is not strong enough to cause synchronization, but they do display variability associated with the rotation of the white dwarf.  Of the at least 15 CVs detected by {\em INTEGRAL}, the vast majority are these relatively rare magnetic systems, with most of those being IPs.

Identification of these unclassified sources requires observations of the source in the optical and/or infrared.  The IBIS imager on {\em INTEGRAL} is unique among hard X-ray/soft $\gamma$-ray detectors in that it is able to locate point sources with an accuracy on the order of arcminutes \citep{gros03}.  To identify a unique optical or infrared counterpart, however, the error circle must be reduced to the level of (sub-)arcseconds.  Thus, from the {\em INTEGRAL} position, soft X-ray telescopes, such as the {\em Chandra X-ray Observatory}, {\em XMM-Newton}, and {\em Swift} observe the region.  This typically allows identification of the optical counterpart, which can then be studied to yield an identification \citep[e.g.,][]{tomsick06}.

 \begin{table*}
\begin{center}
\caption{ Observations: Kitt Peak Optical Spectroscopy\label{tab:obs}}
\footnotesize
\begin{tabular}{cccccc} \hline \hline
IGR Name & Optical Counterpart & RA (J2000) & Dec (J2000) & Start Time (UT) & Exposure Time (s)\\ \hline \hline
J16426+6536 & USNO-B1.0 1555--0172189 & $16^h43^m03^s.99$ & $+65^{\circ}32^{\prime}51~ 2^{\prime\prime}$ & 2008 June 29, 4.0 h & 3600\\
J18214--1318 & USNO-B1.0 0766--0475700 & $18^h21^m19^s.72$ & $-13^{\circ}18^{\prime}38~ 8^{\prime\prime}$ & 2008 June 28, 9.3 h& 3600\\
J18308--1232 & USNO-B1.0 0774--0551687 & $18^h30^m49^s.87$ & $-12^{\circ}32^{\prime}19~ 2^{\prime\prime}$ & 2008 June 30, 8.8 h& 1800\\
J19267+1325 & USNO-B1.0 1033--0440651 & $19^h26^m27^s.00$ & $+13^{\circ}22^{\prime}04~ 4^{\prime\prime}$ & 2008 June 28, 10.6 h& 3 $\times$ 600 \\
J22292+6647 & USNO-B1.0 1567--0242133 & $22^h29^m13^s.90$ & $+66^{\circ}46^{\prime}51~ 9^{\prime\prime}$ & 2008 June 30, 11.2 h& 827 \\ \hline
\end{tabular}
\end{center}
\end{table*}

\section{Observations and Analysis}

We targeted {\em INTEGRAL} sources that had also been observed in soft X-rays, which allowed the position to be narrowed down from the {\em INTEGRAL} error circle (on the order of arcminutes) to less than $10^{\prime\prime}$.  IGR J18214--1318, IGR J19267+1325, and IGR J18308--1232 were observed by the {\em Chandra X-Ray Observatory}, providing positions with 90\% confidence uncertainties of $0.64^{\prime\prime}$ (Tomsick et al. 2008 a, b and in preparation).  IGR J22292+6647 was observed by {\em Swift} \citep{landi07_atel}, which reduced the position uncertainty to $3.6^{\prime\prime}$.  IGR J16426+6536 has one {\em XMM-Newton} source with a $1\sigma$ uncertainty of $8^{\prime\prime}$ \citep{ibarra08a} and two {\em ROSAT} sources within the {\em INTEGRAL} error circle.  Optical/infrared counterparts for IGR J18214--1318 and IGR J19267+1325 were reported with the X-ray observations.  We searched the USNO-B1.0 and 2MASS catalogs for optical and infrared counterparts to the other sources.  IGR J18308--1232 and IGR J22292+6647 both have one optical/infrared counterpart.  The X-ray sources associated with IGR J16426+6536 have three possible optical/infrared counterparts, and we observed the brightest optical source in the {\em XMM-Newton} error circle.  

\begin{figure}
\includegraphics[clip, trim=2.5cm 6.5cm 2.5cm 6.5cm, scale=0.45]{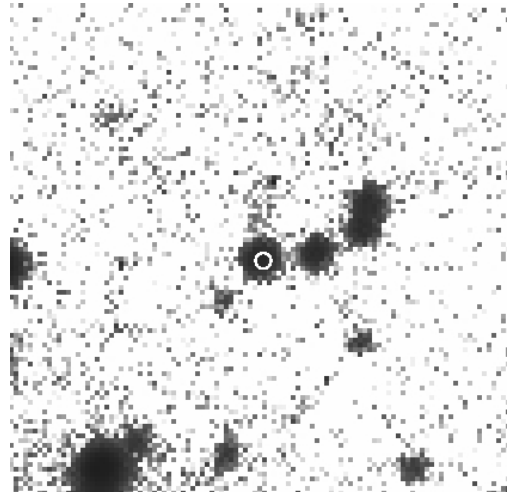}
\caption{I-band image of the field of IGR J18214--1318 with the source ($I\sim17~{\rm mag}$) indicated.  The white circle is the {\em Chandra} error circle of $0.64^{\prime\prime}$.  Three 300s exposures are combined, and the image is $40^{\prime\prime}\times 40^{\prime\prime}$ and oriented so north is up and east is to the left.}
\label{fig:hist}
\end{figure}

Images from the Digitized Sky Survey (DSS) showed a clear counterpart for all sources except for IGR J18214--1318.  A clear image of this source is seen in Figure~\ref{fig:hist}, from  the medium resolution spectrometer TFOSC (${\rm T\ddot{U}B\dot{I}TAK}$ Faint Object Spectrometer and Camera) which is mounted on the Russian-Turkish 1.5 m telescope (RTT150) located at Turkish National Observatory (TUG), Antalya, Turkey.  The camera is equipped with a $2048 \times 2048$, $15~ \mu$m pixel ($0.39^{\prime\prime}$ pixel-1) Fairchild 447BI CCD chip.  We took three 300s observations of the field in B, V, R and I filters on 2008 August 22, and only detected the source in the I band.  After the standard bias and flat correction, we obtained the instrumental I magnitude of the counterpart using DAOPHOT in MIDAS.  We carried out point spread function photometry for the corrected image to obtain the instrumental magnitude, and calibrated it by comparing to the magnitudes of the reference stars, which were obtained from the USNO-B1.0 catalog.

Between 2008 June 28 and June 30 we carried out spectroscopy of these targets with the RC Spectrograph on the 4-meter Mayall Telescope at Kitt Peak National Observatory.  We provide an observing log in Table~\ref{tab:obs}.  We used a slit width of $1.5^{\prime\prime}$, and rotated the slit to the parallactic angle.  The wavelength range covered from $4750~ {\rm \AA}$ to $9500 ~{\rm \AA}$, with a dispersion of $\sim$$3.4{\rm ~ \AA /pixel}$ and a resolution of $13.8 {\rm~ \AA}$.

After applying the flat field correction, subtracting the bias and dark current, and removing cosmic rays, we extracted the spectra with standard procedures in IRAF (Image Reduction and Analysis Facility).  We observed arc lamps for wavelength calibration immediately preceding and following each source.  Checking against background sky lines, we find agreement to within $\sim$$3{\rm~ \AA}$.  We also observed spectrophotometric standard stars Feige 110 and BD $+33^{\circ}~ 2642$ for flux calibration.  Conditions were variable throughout each night, making the flux calibration somewhat uncertain.  We report fluxes using a single observation of Feige 110 for calibration because it falls roughly halfway between the other observations in flux, presumably providing a good indication of the average conditions.  Note that using the other observations can change the flux measurements by up to $\sim$15\%.  All lines and their parameters are reported in Table~\ref{tab:lines}.

\begin{table*}
\caption{Spectral Lines of IGR Sources\label{tab:lines}}
\begin{minipage}{\linewidth}
\begin{center}
\footnotesize
\begin{tabular}{ccccccc} \hline \hline
Line & Quantity\footnote{EW is the equivalent width in ${\rm \AA}$ and line flux is measured in units of $10^{-15} {\rm ~erg~cm^{-2}~s^{-1}}$.  Errors are at the 68\% confidence level and upper limits are at the 90\% confidence level.  Note that the line flux is also subject to the 15\% systematic uncertainty in overall flux level, which is not included here.  The wavelength from the Gaussian fit is $\lambda_{fit}$, and the error is dominated by the $\sim3{\rm~\AA}$ uncertainty in wavelength calibration.}  & J16426+6536 & J18214--1318\footnote{The fitted wavelengths of the Paschen and He {I} absorption lines are $\lambda_{P9}=9234$, $\lambda_{P10}=9016$, $\lambda_{P11}=8865$, $\lambda_{P12}=8753$, $\lambda_{P13}=8669$, $\lambda_{P15}=8545$, $\lambda_{P16}=8501$, $\lambda_{P17}=8469$, $\lambda_{He~ {I}}=8780$.} & J18308--1232 & J19267+1325 & J22292+6647\\ 
     &    &  (AGN)              & (HMXB)          &  (CV)                 &  (CV)                 & (AGN) \\ \hline \hline
${\rm H\alpha}$ & EW & $290 \pm 4^c$ & $<4.3$ & $26 \pm 1$ & $64 \pm 1$ & $760 \pm 20^c$\\
 & Flux & $13.8 \pm 0.2^c$ & $<0.1$ & $7.7\pm 0.2$ & $28.4 \pm 0.2$ & $27.2\pm 0.8$\footnote{${\rm H\alpha}$ is blended with ${\rm [N~II]}$ lines.}\\ 
 & $\lambda_{fit}$  & $8683\pm3^c$ & \ldots & $6562\pm3$ & $6564\pm3$ & $7307\pm3^c$\\ \hline
${\rm H\beta}$ & EW   & $46\pm 2$ & \ldots  & $9.7\pm 1.3$ & $49 \pm 2$ & \ldots\\
 & Flux  & $2.0 \pm 0.1$ & \ldots & $3.1\pm 0.4$ & $15\pm 1$ & \ldots\\
 & $\lambda_{fit}$  & $6430\pm3$ & \ldots & $4860\pm3$ & $4859\pm3$ & \ldots\\ \hline
${\rm H\gamma}$ & EW & $25\pm 3$ & \ldots & \ldots & \ldots & \ldots \\
 & Flux & $1.0\pm 0.1$ & \ldots & \ldots & \ldots & \ldots \\
 & $\lambda_{fit}$  & $5746\pm4$ & \ldots & \ldots & \ldots & \ldots\\ \hline
${\rm [O~III]~\lambda 4959}$ & EW & $7.2 \pm 1.1$ & \ldots & \ldots & \ldots & \ldots \\ 
 & Flux & $0.32\pm 0.05$ & \ldots & \ldots & \ldots & \ldots \\
 & $\lambda_{fit}$ & $6560\pm3$ & \ldots & \ldots & \ldots & \ldots \\ \hline
${\rm[O~III]~\lambda 5007}$ & EW & $32 \pm 1$ & \ldots & \ldots & \ldots & \ldots \\
 & Flux & $1.4\pm 0.1$ & \ldots & \ldots & \ldots & \ldots \\
 & $\lambda_{fit}$ & $6624\pm3$ & \ldots & \ldots & \ldots & \ldots \\ \hline
${\rm O~I~\lambda 7772,7774,7775}$ & EW & \ldots & \ldots & \ldots & $2.2\pm 0.4$ & \ldots \\
 & Flux & \ldots & \ldots & \ldots & $1.2 \pm 0.2$& \ldots \\
 & $\lambda_{fit}$ & \ldots & \ldots & \ldots & $7774\pm3$& \ldots \\ \hline
${\rm He ~I ~\lambda 4921}$& EW & \ldots & \ldots & \ldots & $4\pm 1$& \ldots \\
 & Flux  & \ldots & \ldots & \ldots &$1.1\pm 0.4$& \ldots \\
 & $\lambda_{fit}$  & \ldots & \ldots & \ldots &$4920\pm4$& \ldots \\ \hline
${\rm He ~I ~\lambda 5015}$ & EW & \ldots & \ldots & \ldots & $5\pm 1$& \ldots \\
 & Flux& \ldots & \ldots & \ldots &$1.5\pm 0.3$& \ldots \\
 & $\lambda_{fit}$& \ldots & \ldots & \ldots &$5015\pm3$& \ldots \\ \hline
${\rm He ~I ~\lambda 5876}$ & EW & \ldots & \ldots & $4.6\pm 0.6$& $12\pm 1$& \ldots\\
 & Flux  & \ldots & \ldots & $1.4\pm0.2$ &$5.0\pm 0.2$& \ldots\\
 & $\lambda_{fit}$  & \ldots & \ldots & $5874\pm3$ &$5876\pm3$& \ldots\\ \hline
${\rm He ~I ~\lambda 6678}$ & EW & \ldots & \ldots & $3.8\pm 0.5$ & $8.2\pm 0.4$& \ldots\\
 & Flux & \ldots & \ldots & $1.2\pm 0.2$ &$3.8\pm 0.2$& \ldots \\
 & $\lambda_{fit}$ & \ldots & \ldots & $6677\pm3$ &$6678\pm3$& \ldots \\ \hline
${\rm He~I~\lambda 7065}$ & EW & \ldots & \ldots & $2.9\pm 0.5$ & $6.1\pm 0.4$& \ldots \\
 & Flux & \ldots & \ldots & $0.91\pm 0.15$ & $3.1\pm 0.2$& \ldots \\
 & $\lambda_{fit}$ & \ldots & \ldots & $7063\pm3$ & $7065\pm3$& \ldots \\ \hline
${\rm He~II~\lambda 5412}$ & EW & \ldots & \ldots & \ldots & $5.1\pm 0.9$& \ldots \\
 & Flux & \ldots & \ldots & \ldots & $1.8\pm 0.3$& \ldots \\
 & $\lambda_{fit}$ & \ldots & \ldots & \ldots & $5412\pm3$& \ldots \\ \hline
\end{tabular}
\end{center}
\end{minipage}
\end{table*}

\section{Results}
\subsection{IGR J16426+6536}
IGR J16426+6536 has one {\em XMM-Newton} slew source at $\alpha(J2000)=18^h17^m22^s$, $\delta(J2000)=-25^{\circ}08^{\prime}38^{\prime\prime}$ \citep{ibarra08a} and two {\em ROSAT} sources within the {\em INTEGRAL} error circle.  There is one USNO-B1.0 source within one of the {\em ROSAT} error circles and none in the other.  The {\em XMM-Newton} source has two USNO-B1.0 sources within its $8^{\prime\prime}$ ($1\sigma$) error circle.  We observed the brighter of the two sources (USNO-B1.0 1555--0172189) in the {\em XMM-Newton} error circle. 

\begin{figure}
\subfigure{\includegraphics[scale=0.6]{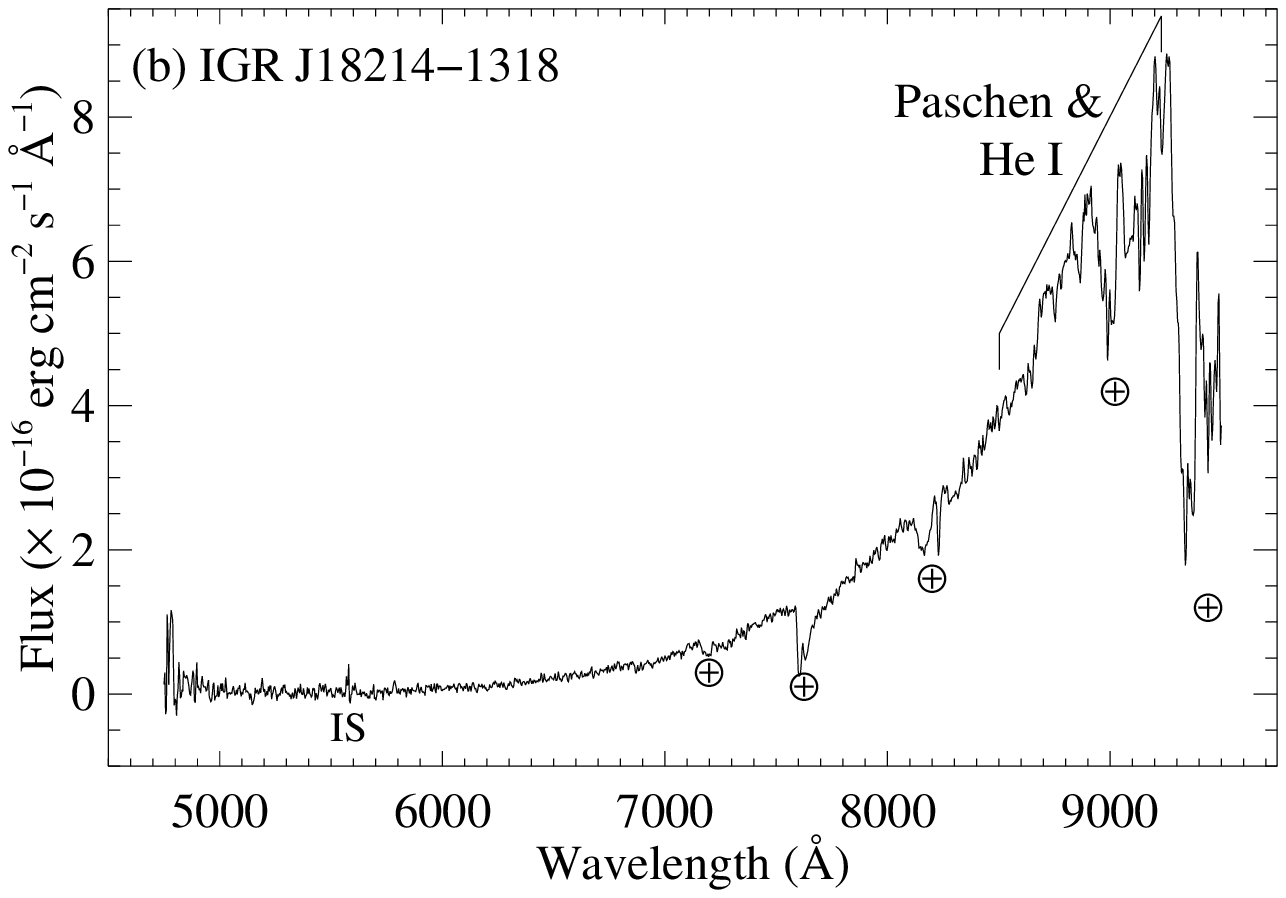}}
\subfigure{\includegraphics[scale=0.6]{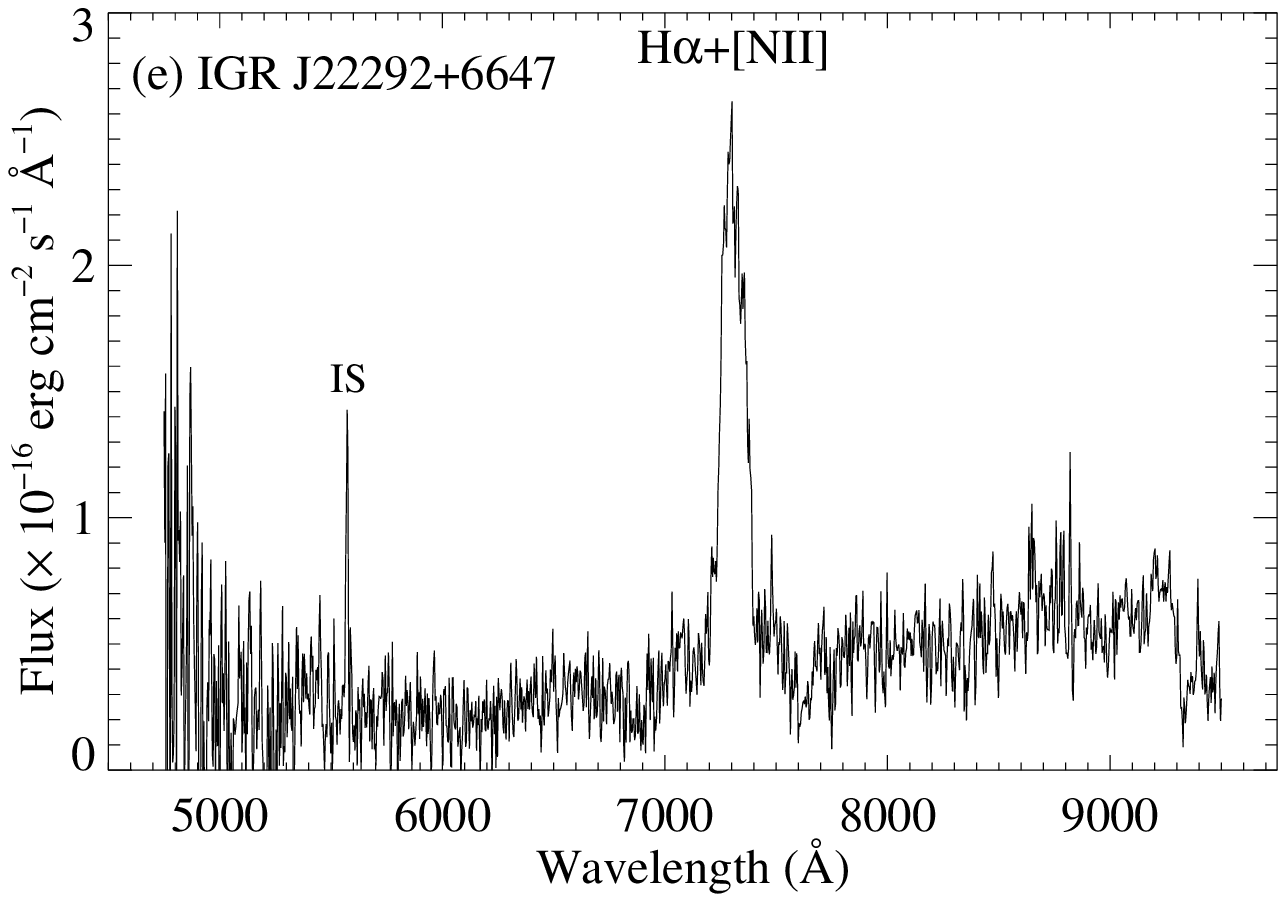}}
\caption{Optical spectra of (a) IGR J16426+6536, an AGN, (b) IGR J18214--1318, an HMXB,
 (c) IGR J18308--1232, a CV, (d) IGR J19267+1325, a magnetic CV, and (e) IGR J22292+6647, an AGN.  The labels ``IS'', 
 ``DIB'', and $\oplus$ indicate interstellar emission lines, diffuse interstellar background, and telluric absorption lines, respectively.}
\label{fig:spectra}
\end{figure}

This spectrum has redshifted emission lines, including broad Balmer lines and narrow forbidden lines (Figure~\ref{fig:spectra}a), indicating that this source is a Seyfert 1 AGN.  This is congruous with its position out of the Galactic plane, with Galactic coordinates $l=96.64^{\circ}, b=+37.65^{\circ}$.  To further classify the source, we use the scheme described in Winkler (1992), namely, the ratio of the flux in $H\beta~ \lambda4861$ to that in [OIII]$\lambda5007$, where these are the laboratory wavelengths.  After dereddening the spectral lines using the Galactic absorption from \cite{schlegel98} along with the extinction law from \cite{cardelli89}, we find $F(4861)/F(5007)=1.5 \pm 0.1$, which makes this a Seyfert 1.5 galaxy.  Calculating the redshift from both $H \beta$ and [OIII]$\lambda5007$ and averaging, we find $z=0.323 \pm 0.001$.  Though we did not observe all possible sources within the {\em INTEGRAL} error circle, the identification of the one we did observe as an AGN allows us to conclude that this is very likely the counterpart of the {\em INTEGRAL} source. We note that \cite{masetti09} came to similar conclusions from their spectroscopic analysis.  Although we identify it as a Seyfert 1.5, while they find that it is a Seyfert 1, there were several months between the observations, and this level of variability is not unheard-of \citep[e.g.,][]{tran92}.

With a redshift of $z=0.323$, this source is at a luminosity distance of $D_{\rm L}=1690~{\rm Mpc}$, using $H_0=70{\rm ~km~s^{-1}~Mpc^{-1}}$, $\Omega_{\rm M}=0.3$, and $\Omega_{\Lambda}=0.7$.  We calculate the mass of the supermassive black hole using the relations described in \cite{wu04} and \cite{kaspi00}.  To use the relations, we recalculate the luminosity distance with the cosmological parameters used in those papers, $H_0=75{\rm ~km~s^{-1}~Mpc^{-1}}$, $\Omega_{\rm M}=1$, and $\Omega_{\Lambda}=0$.  This gives $D_{\rm L}=1380 {\rm ~Mpc}$, from which we find the luminosity in $H\beta$ (dereddened as described above), $L_{\rm H\beta}=4.9 \times 10^{41} {\rm ~erg~s^{-1}}$.  Combining this with the rest-frame FWHM velocity of the broad line region (from $H\beta$), $v_{\rm FWHM}=2000 {\rm ~km~s^{-1}}$, gives a black hole mass of $M_{\rm BH}=9 \times 10^6 ~\Msun$.  
\begin{figure}
\subfigure{\includegraphics[scale=0.6]{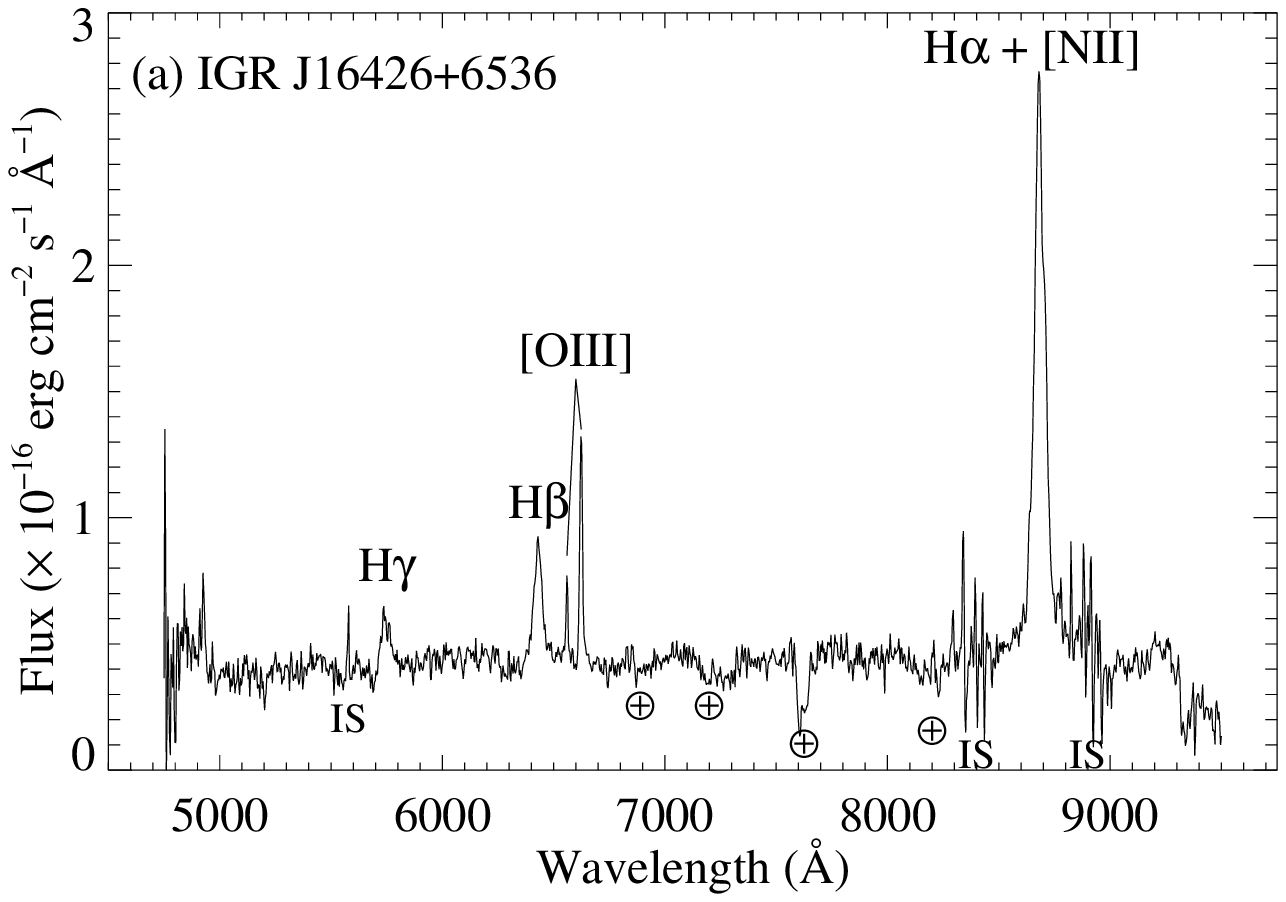}}
\subfigure{\includegraphics[scale=0.6]{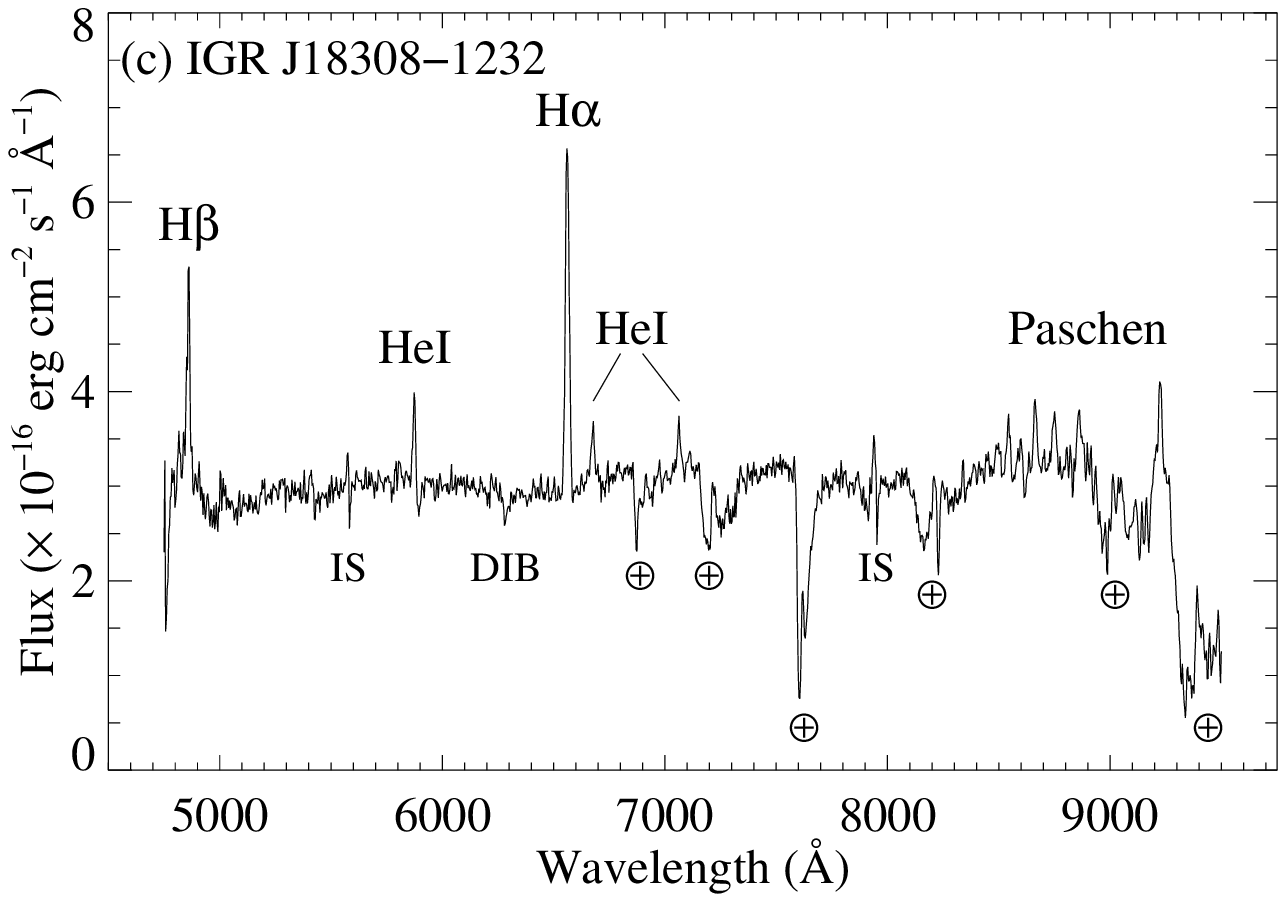}}
\subfigure{\includegraphics[scale=0.6]{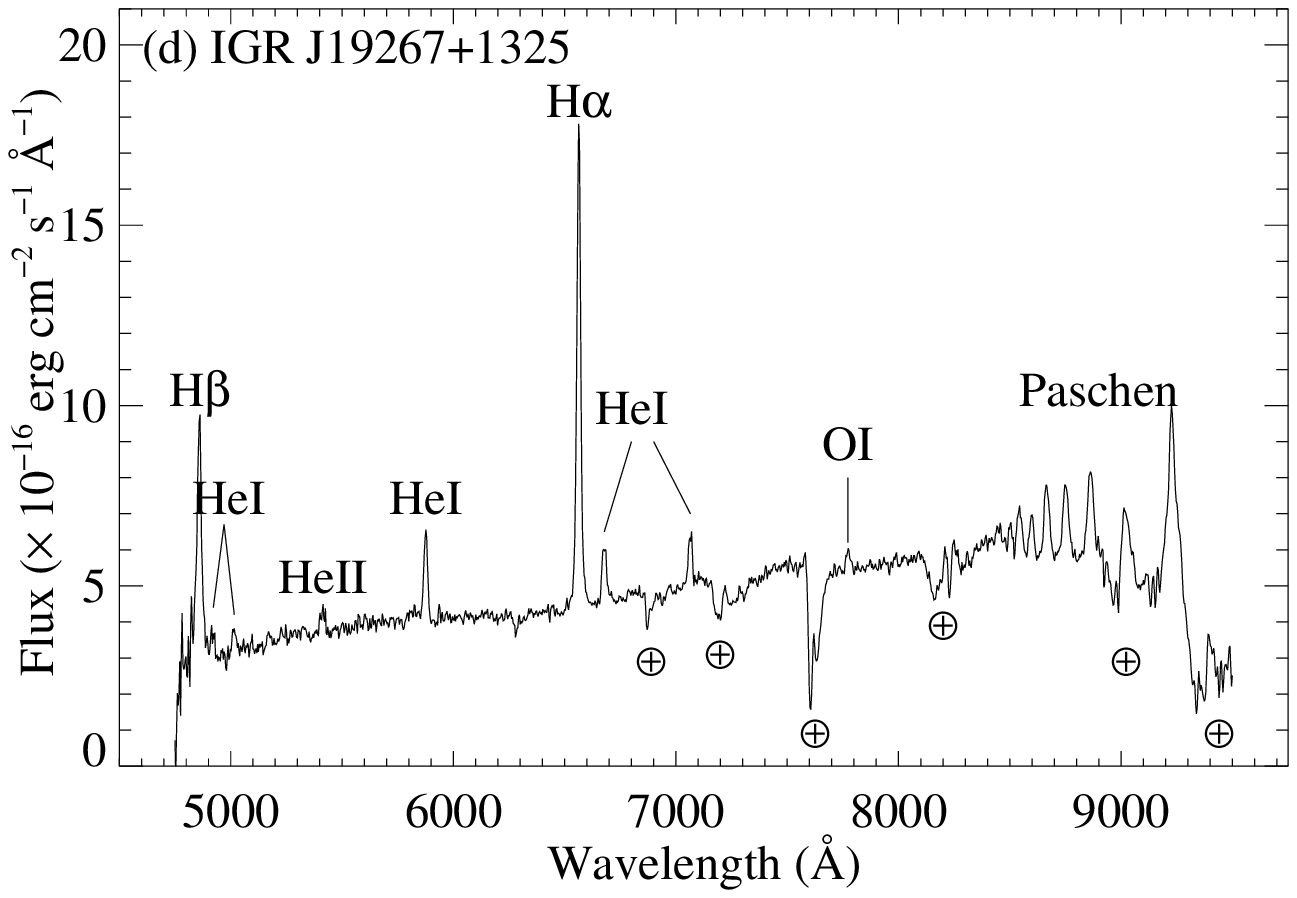}}
\end{figure}

\subsection{IGR J18214--1318}
IGR J18214--1318 was observed by {\em Chandra}, which found one source at $\alpha(J2000)=18^h21^m19^s.76$, $\delta(J2000)=-13^{\circ}18^{\prime}38~9^{\prime\prime}$, coinciding with USNO-B1.0 0766--0475700.  The X-ray spectrum yields a column density of $N_{\rm H}=(11.7^{+3.0}_{-2.7})\times 10^{22}{\rm ~cm^{-2}}$ and  $\Gamma=0.7^{+0.6}_{-0.5}$.  With a Galactic hydrogen column density of $N_{\rm H}=2.4 \times 10^{22} {\rm ~cm^{-2}}$, this implies a local absorption of $N_{\rm H}=(7-15) \times 10^{22} {\rm ~cm^{-2}}$, which was suggested to be from the wind of a high-mass star \citep{tomsick08}.  This source was also observed by {\em Swift}, which found a position and photon index consistent with that obtained by {\em Chandra}, but a lower hydrogen column density of $N_{\rm H}=(3.5^{+0.8}_{-0.5})\times 10^{22}{\rm ~cm^{-2}}$, indicating that $N_{\rm H}$ is variable in this source \citep{rodriguez09}.

We find that the optical spectrum has a very reddened continuum (Figure~\ref{fig:spectra}b).  From this, we infer that the source is relatively distant, since significant reddening would be expected for a distant source in the Galactic plane ($l=17.69^{\circ}, b=+0.48^{\circ}$).  Convolving the spectrum with the filters from \cite{bessell90}, we find $V=22.2\pm 0.2$, $R=19.3\pm 0.2$, $I=16.6\pm 0.2$, where the errors are from the 15\% systematic uncertainty in flux.  We note that the I band magnitude from the photometric observation made at TUG, $I=16.92\pm0.17$, agrees within the errors.  We first consider the possibility that the source is a CV.  Combining the apparent magnitudes with typical absolute magnitudes for CVs ($M_V\sim 9$ and $(V-R)_0\sim 0$ \citep{masetti09}) and a range of extinctions from $A_V=5-15 {\rm ~mag}$ results in distances that range from $d=4 - 440 {\rm~pc}$.  This distance range is far too close to account for the observed amount of extinction (assuming the optical extinction is interstellar).  Furthermore, the spectrum lacks emission lines, while a CV would be expected to have strong emission lines.  These arguments allow us to rule out the possibility that the source is a CV.  Applying the same procedure to check for the possibility of an LMXB, we combined the apparent magnitude, the typical absolute magnitude for an LMXB ($M_V\sim 0$ and $(V-R)_0\sim 0$ \citep{vpmc95}), and the same range of extinctions.  In this case, the distances range from $d=300 - 28000 {\rm~pc}$, which are not inconsistent with the amount of extinction, but the source has a hard X-ray spectrum which would be very unusual for an LMXB \citep{muno04}.  

\begin{figure}
\includegraphics[clip, trim=5cm 4cm 5cm 5cm, scale=1]{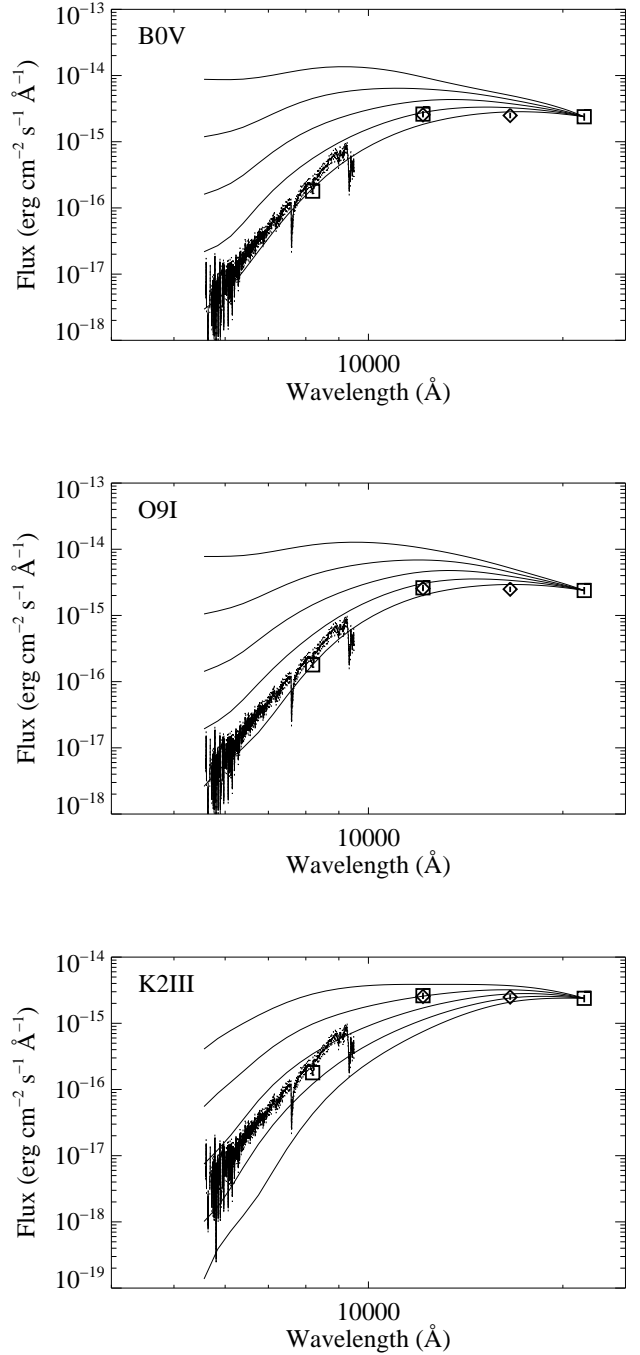}
\caption{The smooth lines are spectral shapes of B0V, O9I, and K2III stars interpolated from published colors with varying degrees of extincion.  Starting from the top line, the spectra are plotted with $A_{\rm V}=5, 7.5, 10, 12.5,{\rm and}~15$.  All are normalized to have the same K magnitude.  The squares indicate the DENIS magnitudes, and the diamonds indicate the 2MASS magnitudes of the optical counterpart of IGR J18214--1318.  The spectrum taken at Kitt Peak is plotted with dotted lines overlayed to indicate the upper and lower bounds due to the 15\% uncertainty in flux calibration.}
\label{fig:hmxbColors}
\end{figure}

\begin{figure}
\includegraphics[clip, trim=1cm 0cm 0cm 0cm, scale=0.65]{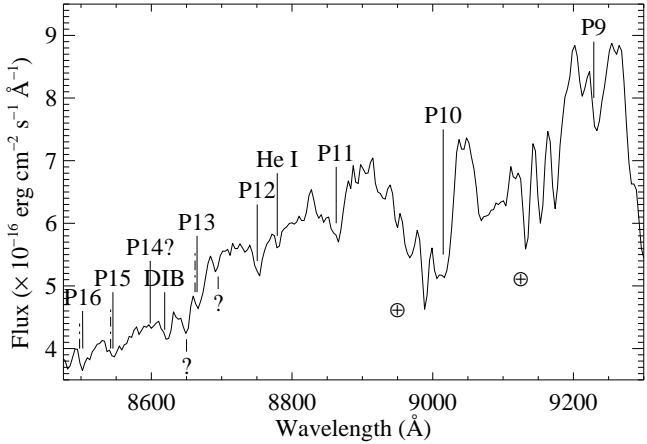}
\caption{Close up of the red portion of the spectrum for IGR J18214--1318.  The dot-dashed lines indicate where the Ca II triplet would be located.  The label ``DIB'' indicates the diffuse interstellar background.}
\label{fig:zoom18214}
\end{figure}

These considerations allow us to conclude that it must be either an HMXB or a symbiotic star system.  In Figure~\ref{fig:hmxbColors}, spectral shapes typical of both are interpolated from the colors in \cite{ducati01} and plotted with extinctions of $A_V=5,7.5,10,12.5,{\rm and}~ 15$, using the extinction relations from \cite{cardelli89}.  The infrared magnitudes from DENIS and 2MASS (reported in Tomsick et al. 2008a) along with the optical spectrum for the source are overlayed.  These show that the data are consistent with a B0V star or an O9I star with extinction of $A_V=12.5-15$.  The model K2III star spectra provide a worse fit to the data.  The J-band magnitude requires an extinction of $A_V\sim7.5$, while the optical data are consistent with $A_V=10-12.5$.  Further evidence supporting an HMXB is shown in Figure~\ref{fig:zoom18214}, which is a close up of the near-infrared region of the spectrum.  Several Paschen absorption lines and one neutral Helium line are observed (although P14 is not observed).  These lines are indicative of an O or B star \citep[e.g.,][]{co06,mt99}.  As a check, we indicate the positions of the Ca {\sc II} triplet with the dot-dashed lines, and confirm that the absorption is more consistent with Paschen lines.  In addition, other Paschen lines (P12, P11, P10, P9) are clearly detected.  We also note that the upper limit on the equivalent width of $H\alpha$ is consistent with $H\alpha$ equivalent widths of OB stars \citep{leitherer88}.  Thus, we firmly identify the optical source as a high mass star.

The Galactic hydrogen column density combined with the relationship between visual extinction and hydrogen column density detailed in \cite{ps95} ($A_V=0.56N_H[10^{21} {\rm cm^{-2}}]+0.23$) gives a visual extinction of $A_V=13.7$, which is consistent with the extinction inferred from colors for stars of type B0V and O9I.  Considering a range of spectral types and using absolute magnitudes from \cite{cox00} with $A_V=13.7$, we find distances that range from $d=3-7~{\rm kpc}$ for B0V to O5V stars and $d=9-10~{\rm kpc}$ for A0I to O9I stars, giving an overall distance range of $d=3-10~{\rm kpc}$.  However, using the general relation of $2~A_V/{\rm kpc}$ \citep{tielens05}, the lower limit of visual extinction ($A_V=12.5$) implies a distance of at least 6 kpc, suggesting that the most likely distance range is $d=6-10~{\rm kpc}$.  In addition, the observed variation in hydrogen column density argues in favor of a supergiant (and, therefore, a distance near the upper end of the range), since HMXBs with supergiant companions are known to show variations in column density \citep{prat08}.

\subsection{IGR J18308--1232}
IGR J18308--1232 was observed by {\em Chandra}, which found one source at  $\alpha(J2000)=18^h30^m49^s.94$, $\delta(J2000)=-12^{\circ}32^{\prime}19 ~1^{\prime\prime}$, coinciding with USNO-B1.0 0774--0551687. The X-ray spectrum gives a hydrogen column density of $N_{\rm H}=(3.1^{+2.9}_{-2.4})\times 10^{21} {\rm ~cm^{-2}}$ and $\Gamma=0.41^{+0.25}_{-0.30}$, where the errors are at the 90\% confidence level (Tomsick et al. in prep.).  Note that this source is referred to as IGR J18307--1232 in \cite{ibarra08b}, where they report an {\em XMM-Newton} slew-survey counterpart at $\alpha(J2000)=18^h30^m49^s.6$, $\delta(J2000)=-12^{\circ}32^{\prime}18^{\prime\prime}$ with a $1\sigma$ uncertainty of $8^{\prime\prime}$, consistent with the {\em Chandra} position.

The optical spectrum shows Balmer, Paschen, and HeI lines in emission (Figure~\ref{fig:spectra}c), all consistent with $z=0$.  These lines are typical of both CVs \citep{warner95} and LMXBs \citep{vpmc95}, but the hard X-ray spectrum is typical for a CV, and, as noted above, is not expected for an LMXB.  Using the hydrogen column density and the aforementioned relation between column density and visual extinction, we find $A_V=2.0^{+1.9}_{-1.6}$.  Convolving the spectrum as described in the previous section gives $V=17.7 \pm0.2$, $R=17.1 \pm0.2$, and $I=16.4 \pm 0.2$.  Combining the magnitude and visual extinction with the typical values for CV absolute magnitudes listed above, we find $d=220^{+160}_{-190} {\rm ~pc}$.  \cite{masetti09} also identify this as a CV.  They find a larger distance, although it agrees within the uncertainties, of $d=320 {\rm ~pc}$ as a result of using USNO-A2.0 magnitudes (which are fainter for this source) and a different method of calculating extinction (which resulted in lower extinction).

\subsection{IGR J19267+1325}
IGR J19267+1325 was observed by {\em Chandra}, which found one source at $\alpha(J2000)=19^h26^m26^s.99$, $\delta(J2000)=+13^{\circ}22^{\prime}05~ 1^{\prime\prime}$, coinciding with USNO-B1.0 1033--044065.  Although the source for IGR J19267+1325 falls just outside of the $90\%$ confidence {\em INTEGRAL} error circle, it was the only bright X-ray source {\em Chandra} found in the area.  Fitting the X-ray spectrum gives $N_{\rm H}=(2.1\pm 0.9) \times 10^{21} {\rm ~cm^{-2}}$ and $\Gamma=0.68\pm 0.13$ \citep{tomsick08_atel}.
  
The optical spectrum shows Balmer, Paschen, HeI and HeII emission lines at $z=0$ (Figure~\ref{fig:spectra}d).  These lines, along with the hard X-ray spectrum, suggest a CV nature \citep[e.g.,][]{masetti09}.  Furthermore, the HeII line and large equivalent width of $H\beta$ ($49 \pm 2 {\rm \AA}$) suggest a magnetic CV \citep{silber92}\footnote{See http://asd.gsfc.nasa.gov/Koji.Mukai/iphome/issues/heii.html}.  \cite{steeghs08_atel} and \cite{masetti09} also identify this source as a probable magnetic CV.  The equivalent widths reported by \cite{masetti09} match our findings ($EW_{H\alpha}=64\pm 1{\rm ~\AA}$), while \cite{steeghs08_atel} find an equivalent width of $H\alpha$ equal to $120 {\rm ~\AA}$.  Based on two measurements from the INT Photometric H-Alpha Survey (IPHAS), they find moderate short term variability in $H\alpha$ of ~0.1 mag, which is not large enough to explain the difference in equivalent widths found, though there could be a larger maximum variability.  Based on the detection of a periodicity in {\em Swift-XRT} data, \cite{evans08} confirm that it is a magnetic CV, identifying it as an intermediate polar CV.  Using the hydrogen column density and the relation between visual extinction and hydrogen column density mentioned previously, we find $A_V=1.4\pm 0.7$.  Convolving the spectrum as described above, we find magnitudes of $V=17.4 \pm0.2$, $R=16.7 \pm0.2$, and $I=15.8 \pm 0.2$.  We combine the magnitude and visual extinction with the typical values of CVs above to find $d=250\pm 80{\rm ~pc}$.  This distance is lower than that found by \cite{masetti09}, $d=580 {\rm ~pc}$, for the same reasons as for IGR J18308--1232.

\subsection{IGR J22292+6647}
IGR J22292+6647 was observed by {\em Swift}, which found a source at $\alpha(J2000)=22^h29^m13^s.5$, $\delta(J2000)=+66^{\circ}46^{\prime}51~ 8^{\prime\prime}$, coinciding with USNO-B1.0 1567--0242133.  There is a radio source at this location (87GB 222741.2+663124), described as an asymmetric double \citep{gc91}, which suggests this may be an AGN \citep{landi07_atel, landi09}.  This is congruous with its position out of the Galactic plane, with Galactic coordinates $l=109.56^{\circ}, b=+7.69^{\circ}.$

The optical spectrum has a flat continuum dominated by one strong and broad emission line (Figure~\ref{fig:spectra}e).  If we assume that the strong feature is $H\alpha$, we find a redshift of $z=0.113 \pm 0.001$.  The broad hydrogen line indicates this is a Seyfert 1 AGN.

\section{Discussion and Conclusions}
We have firm identifications of five IGR sources.  IGR  J18308--1232 and IGR J19267+1325 are CVs.  They have distances on the order of a couple hundred parsecs, which is typical of other CVs detected by {\em INTEGRAL} \citep[e.g.,][]{barlow06,masetti06}.  The column densities from X-ray observations imply extinctions that are rather large given the distances we find.  For example, IGR J19267+1325 is found to have 1.4 magnitudes of extinction at a distance of 250 pc, while typical average extinction is $A_V=2{\rm~mag~kpc^{-1}}$ \citep{tielens05}.  However, more reasonable distances and extinctions are feasible within the uncertainties.  This could also mean that some of the X-ray absorption is intrinsic to the source.  The identification of  IGR J19267+1325 as a magnetic CV is not surprising, and, in fact, IGR  J18308--1232 was also identified as a magnetic CV by \cite{masetti09}.  As a result of the hard X-rays emitted by magnetic systems, the majority of CVs detected by {\em INTEGRAL} have been magnetic \citep{barlow06} even though such systems only comprise $\sim10\%$ of the CV population as a whole.  {\em INTEGRAL} observations of CVs, therefore, unambiguously show that the magnetic field plays an important role in the hard X-ray emission of these systems.

We conclude that IGR J18214--1318 is an obscured HMXB.  It has an extremely reddened continuum, indicating much absorption along the line of sight, which implies a fairly large distance.  It has absorption lines that are typical of a high mass star, and the observed colors and flux are consistent with an extincted high mass main sequence or supergiant star (and the variable hydrogen column density may argue for the latter).  In addition, the X-ray spectrum shows significant local absorption, which could come from a wind from a high mass star.  An infrared spectrum may be one way to further classify the companion star.  The infrared suffers less extinction than the optical, and more features would be observable.  A higher resolution spectrum would also be valuable in identifying the luminosity and exact spectral class.

It is instructive to compare the X-ray luminosities for the above sources to typical HMXBs and CVs.  In the {\em Chandra} band (0.3--10 keV), IGR J18214--1318 has an unabsorbed 0.3--10 keV flux of $(60^{+317}_{-35}) \times 10^{-12} {\rm ~erg~cm^{-2}~s^{-1}}$ \citep{tomsick08}.  Taking a distance of $8~{\rm kpc}$, this results in a luminosity of $L_{\rm 0.3-10~keV}=5 \times 10^{36} {\rm ~erg~s^{-1}}$.  This is consistent with the average properties of neutron star HMXBs given in \cite{vpmc95}.  While the hard X-ray spectrum and luminosity argue for a neutron star, detection of X-ray pulsations would be necessary to rule out  a black hole.  IGR  J18308--1232 has $F_{\rm 0.3-10~keV}=(2.3^{+0.8}_{-0.4}) \times 10^{-11} {\rm ~erg~cm^{-2}~s^{-1}}$ (Tomsick et al. in prep.), which implies a luminosity of $L_{\rm 0.3-10~keV}=1.3 \times 10^{32}{\rm ~erg~s^{-1}}$ with our calculated distance (220 pc).  The  {\em INTEGRAL} flux from $20-100 {\rm ~keV}$ of $F_{\rm 20-40~keV}=0.8 \pm 0.1  {\rm~mCrab}$ and $F_{\rm 40-100~keV}=1.0 \pm 0.2 {\rm~mCrab}$ \citep{bird07} implies a luminosity of $L_{\rm 20-100~keV}=9 \times 10^{31} {\rm ~erg~s^{-1}}$.   We compare this luminosity to known intermediate polar CVs in \cite{suleimanov05} over the energy range $0.1-100 {\rm ~keV}$, assuming the flux from the range where we lack data is of the same order as that from the {\em Chandra} range (a reasonable assumption, given the spectral shapes in the above paper).  Our luminosity falls in the range of known IP CVs, although it is on the low end.  IGR J19267+1325 is a similar case.  Its flux in the {\em Chandra} band is $F_{\rm 0.3-10~keV}=(8.9 \pm 0.6) \times 10^{-12} {\rm ~erg~cm^{-2}~s^{-1}}$ \citep{tomsick08_atel}, which implies a luminosity of $L_{\rm 0.3-10~keV}=6.7 \times 10^{32}{\rm ~erg~s^{-1}}$ with our calculated distance (250 pc).  The  {\em INTEGRAL} flux of $F_{\rm 20-40~keV}=0.7 \pm 0.1  {\rm~mCrab}$ and $F_{\rm 40-100~keV}=0.6 \pm 0.2 {\rm~mCrab}$ \citep{bird07} implies a luminosity of $L_{\rm 20-100~keV}=8 \times 10^{31} {\rm ~erg~s^{-1}}$.   Again, this results in a total luminosity on the low end of the range of known CV luminosities.  These low values could be an indication that the distances lie at the high end of our error range.  

\begin{figure}
\includegraphics[scale=0.6, clip, trim=1.2cm 0cm 0cm 0cm]{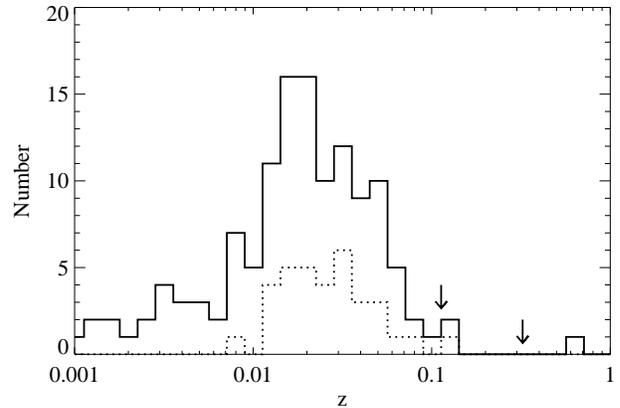}
\caption{The solid line is the number of Seyfert galaxies detected by {\em INTEGRAL} and the dotted line is number of IGR Seyfert galaxies as a function of redshift \citep{bodaghee07}.  The arrows indicate IGR J22292+6647, with $z=0.113$, and IGR J16426+6536, with $z=0.323$.}
\label{fig:agnHist}
\end{figure}

IGR J16426+6536 and IGR J22292+6647 are Seyfert 1 AGN, and IGR J16426+6536 is further classified as a Seyfert 1.5 AGN.  The 127 {\em INTEGRAL}-detected Seyferts (which have an average redshift $z=0.033$) and the 34 IGR Seyferts (with average redshift $z=0.035$) are plotted in Figure~\ref{fig:agnHist}.  With redshifts of $z=0.323$ and $z=0.113$, respectively, IGR J16426+6536 and IGR J22292+6647 are considerably more distant than average.  In fact, IGR J16426+6536 is an interesting case since it is the highest redshift IGR Seyfert, though there is one {\em INTEGRAL}-detected Seyfert, PKS 0637--752, with a higher redshift of $z=0.651$.

\acknowledgements
J. A. T. and S. C. B. acknowledge partial support from {\em Chandra} awards G07-8050X and G08-9055X issued by the {\em Chandra X-Ray Observatory} Center, which is operated by the Smithsonian Astrophysical Observatory for and on behalf of the National Aeronautics and Space Administration (NASA), under contract NAS8-03060.  S. C. B. is partially supported by a UC Berkeley Chancellor's Fellowship and an SSL Summer Science Fellowship.  E. K.  acknowledges partial support from TUB\.ITAK, and Turkish National Academy of Sciences Young and Successful Scientist award (GEB\.IP), and European Commission FP 6 Marie Curie Transfer of Knowledge Project ASTRONS, MTKD-2006-42722.  We thank Hal Halbadel and Daryl Willmarth for their help at Kitt Peak and Jules Halpern for sharing optical images.  This publication makes use of data products from the Two Micron All Sky Survey, which is a joint project of the University of Massachusetts and the Infrared Processing and Analysis Center/California Institute of Technology, funded by NASA and the National Science Foundation.  This research also makes use of the USNOFS Image and Catalogue Archive operated by The United States Naval Observatory, Flagstaff Station and the Deep Near Infrared Survey of the Southern Sky (DENIS).  We also used the Digitized Sky Survey, which was produced at the Space Telescope Science Institute under U.S. Government grant NAG W-2166. The images of these surveys are based on photographic data obtained using the Oschin Schmidt Telescope on Palomar Mountain and the UK Schmidt Telescope.  We thank the anonymous referee for helpful suggestions.


\end{document}